\documentclass[twocolumn,twoside,slac_two]{revtex4}
\usepackage{graphicx}
\usepackage{fancyhdr}
\begin{document}

\title{FPCP 2012 Summary Talk on Experiments} 

\author{Jeffrey A. Appel}
\affiliation{Fermilab, Batavia, IL 60510, USA}

\begin{abstract}
In over forty presentations on experiments at the 2012 conference on Flavor Physics and CP Violation (FPCP 2012), there was an abundance of beautiful 
and significant results. This summary of these experiment presentations begins 
with a reminder of the context in which the measurements have been made and 
the motivations for making the specific measurements reported at the symposium. 
Given the number and breadth of physics topics covered at the meeting, this 
review covers only a limited set of highlights, sort of a traveler's set of 
souvenir postcards of favorite slides. The selected slides are grouped into 
eight overlapping categories as an aid to flipping through the postcards and 
being reminded of the high points of the conference. Finally, there are some 
summarizing comments about how the experiment results presented here compare 
to expectations and what we may hope for the future.     
\end{abstract}

\maketitle

\thispagestyle{fancy}


\section{Introduction}

Before turning to the experiment presentations at FPCP 2012, it may be useful 
to review the broader context of the FPCP conference and of the measurements. 
We often say that particle physics is the study of matter, energy, space, and 
time. What do we really want to know? 

For matter, we want to know:
\begin{itemize}\itemsep0pt
\item Why is the universe so dominantly matter; why is there so little 
antimatter around?
\item Why is matter made of quarks and leptons, antiquarks and antileptons?
\item Why are these constituents spin $\frac{1}{2}$ particles?
\item Why do quarks and leptons come in three generations? (And, are there only 
three?)
\item Why do the quark generations have such different masses?
\item Why are neutrino masses so small; and why different by generation? Are 
neutrinos Dirac or Majorana particles?
\item Why do quarks come in three colors?
\end{itemize}
For energy: 
\begin{itemize}\itemsep0pt
\item Why is so little of the energy density in the universe composed of the 
mass that I just listed? 
\item What is the dark matter we claim is the rest of the matter?
\item Why are the force carriers spin 1; not spin 0 for example? Why are 
gravitons spin 2?
\item Why are the strong and electroweak forces flavor independent?
\item Why aren't all interactions flavor independent?
\end{itemize}
And, for space and time:
\begin{itemize}\itemsep0pt
\item Why are there three obvious spatial dimensions?  Are there more?
\item Why is the expansion of the universe accelerating?  Or, do we not 
understand gravity/space?
\end{itemize}

As an aside, note the preponderance of the number three in my list of questions:  
three dimensions, three generations, and three colors. Are all these trinities 
related?  Are any of them related?

Sometimes we have ideas about possible keys to unlocking the answers to some of 
these questions:
\begin{itemize}\itemsep0pt
\item Some kind of substructure to explain the pattern of quarks and 
 leptons and generations we see.
\item The Higgs mechanism as what nature has chosen for ElectroWeak Symmetry
 breaking? Do the masses of the $W$ and $Z$ come from the same mechanism as 
 that for quark masses.
\item A seesaw mechanism involving very massive right-handed partners as the 
 source of the very light neutrino masses.
\item The neutrino sector as the source of the matter/antimatter asymmetry 
 of the visible universe. We know now that the asymmetry is not from the CKM 
 matrix in the quark sector (inadequate by 8-10 orders of magnitude). 
\end{itemize} 

Not all these ideas are directly testable. Those that are, we are working hard 
to test. We are doing it by going to higher energies at the LHC and by making 
more and more precise measurements in the flavor sector. The latter is the focus 
of our FPCP conferences, of course. 

I am reminded of the story of the drunk and the lamp post. Maybe this is only 
well known in the West, so indulge me if I tell it here. A drunk has lost his 
keys and is spending a long time looking for them under a lamp post. When asked 
why he is still looking there, he says that that is where there is enough light 
to see them!  
We need to be careful not to make the same mistake. Certainly, look where we 
think we may find the answers because we have good models to test. However, in 
parallel, probe as deeply as we can where we don't have such light to guide us.
We have vast new data sets, and we need to check for the unexpected, too. As a 
trivial example, in studying decays of particles with heavy flavor to 
$h^{\pm} \ell^+ \ell^-$, look for $h^- \ell^+ \ell^+$, etc. as well.  No matter 
what our theorist friends tell us about where the answers lie, we have already 
seen the preferred space of minimal SUSY disappear. And, the space for the 
Higgs to hide is also closing down. Sorry, I am supposed to say that we are 
closing in on the Higgs at about 125 $GeV/c^2$! (I note in this post conference 
write-up the announcements from the LHC and Tevatron after the FPCP 2012 
conference indicating the possible discovery of a Standard-Model Higgs 
boson \cite{ATLAS-Higgs, CMS-Higgs}!)

\section{Results shown at FPCP 2012}

This brings us to the beautiful results shown at FPCP 2012, many results new 
since the last meeting, some shown here publicly for the first time. Since I 
cannot include all my favorite slides in the write-up of the talk, I can only 
refer the reader to the slides for this summary available from the Proceedings 
link on the conference web site \cite{FPCP} as an accompaniment to reading this 
article. As you look over the slides I have selected from the meeting, you will 
see a personal selection.  Afterall, choosing post cards when you are on travel 
is a personal matter. So, I show some of my favorite postcards from FPCP 2012, 
those I found especially pretty or revealing. 

I won't repeat all the excellent explanations of the results. I could not do as 
well as we have heard from the presenters themselves. These explanations are 
available in slides from the other talks, also available at the conference web 
site, or in the individual write-ups from the presenters, also available on 
links from the conference web site.

I would certainly choose a postcard each to remind me of the conference site, 
the Chinese opera we saw, and the Bao Gong Memorial Park. However, there are 
also lots of postcards to select with pretty plots and physics content.

My souvenir postcards are organized in eight categories:
\begin{itemize}\itemsep0pt
\item Standard-Model confirmations
\item Significant reductions in uncertainty
\item New-Physics space ruled out
\item Tension with the Standard Model
\item New signals and structures
\item Hints of new physics
\item New techniques and looking beyond the lamp post
\item Postcards of the future
\end{itemize}
Remember, these are just souvenir post cards, visually impressive views of 
the various physics topics presented, not part of a guide book. See the 
individual talks for details at the level of a guide book. It should also be 
obvious that most measurements that I reference could be listed for more than 
one category.

\subsection{Standard-Model confirmations}

New measurements of the speed of neutrinos have confirmed that the earlier 
measurement of neutrino speed by the OPERA experiment 
(now revised also by OPERA itself) 
was wrong. Neutrinos are not superluminal. They do not travel faster than the 
speed of light. In his talk, Andrew Cohen noted \cite{cohen} that we should have 
realized this from known physics, in particular the lack of radiation by 
high-energy neutrinos from very far away, for example. The OPERA result did 
motivate serious thinking about the issue, and generated calculations of new, 
very sensitive limits on the violation of Lorentz invariance by neutrinos.  

Andrzej Bozek showed Belle's result on the rate of $D_S \rightarrow \tau \nu$, 
consistent with lepton universality \cite{bozek}. There is also continuing 
progress in measurements of $CP$ violation in the quark sector. Giovanni 
Marchiori showed BaBar's first three-sigma evidence of $CP$ violation in the 
decay of the $B$ to three $K_S^0$'s from an analysis of the time-dependence of 
the decays \cite{marchiori}; and we saw the first three-sigma evidence of $CP$ 
violation in $B_S$ decays from LHCb as shown by Irina Nasteva \cite{nasteva}. 
Also, note the evidence shown by Yuehong Xie that the heavier $B_S$ lives 
longer than its lighter sister (also from LHCb) \cite{xie}. 

Finally, the window for a Standard-Model Higgs continues to be better defined
with the more precise measurements of the $W$ mass by CDF and DZero at the 
Tevatron \cite{charlton}. The continued reduction in the ``oval of uncertainty" 
in the Higgs mass from ever more precise top-quark and $W$ mass measurements 
at the Tevatron (consistent with a light Higgs as predicted from 
electroweak measurements and possibly observed as announced soon after the 
FPCP meeting \cite{ATLAS-Higgs, CMS-Higgs}).

\subsection{Significant reductions in uncertainty}

Perhaps the most startling reduction in a measurement uncertainty has come 
with the surprisingly-quick measurement of $\theta_{13}$ of the neutrino-mixing 
matrix parameterization as shown in the talks of Werner Rodejohann, Jianglai 
Liu, and Phillip Litchfield \cite{rodejohann, liu, litchfield}. The value 
measured, first and best so far by the Daya Bay experiment, is near the 
previous upper limit on this parameter. The optimists were right in this case.

There is also the improvement in the uncertainties in the parameters of the 
so-called unitarity triangles of the CKM quark-mixing matrix as shown in 
individual presentations and summarized in the talk by Sebastien 
Descotes-Genon \cite{genon}.
 
Also impressive are the measurements of the properties of the $h_C$ by 
BESIII shown by Guangshun Huang \cite{huang} and improvements in the now-lower 
value of $y_{CP}$ in $D^0$ decays, including the new measurements from BaBar 
and Belle, reported by Chunhul Chen \cite{c-chen}. 

Finally, I note separately the improvements in $B_S$ mixing parameters, 
including the value of $\phi_S$ reported by Sebastien, Fabrizio Ruffini, 
and Yuehong Xie \cite{genon, ruffini, xie}, progress in reducing the 
semileptonic decay uncertainties, e.g., in bottom decays as reported by Vera 
Luth \cite{luth}, and in charm form factors from BESIII as reported by Jonas 
Rademacker \cite{rademacker}.

\subsection{New physics space ruled out} 

The parameter space for physics beyond the Standard Model is multidimensional.
We have become used to presentations of limits when models are reduced to two 
relevant parameters, whether we are talking about dark matter in terms of cross 
sections and particle mass or SUSY models at the selected internal-parameter 
level. Some of the slides presented show measurement limits for physical 
parameters with a range of model-possibility predictions of the parameters 
as generated by Monte Carlo techniques to give a sense of how effective the 
measurements are in restricting the range of model parameters. Other 
limits come directly from two-dimensional plots of possible model parameters 
with sections of the space ruled out by the measurements.  Some of the plots 
are quite colorful and artistic in appearance, as well as providing physics 
insight. 

Some of my favorite plots come from the presentation on dark-matter search 
limits by Xinchou Lou \cite{lou} both for generic dark matter and for a 
possible dark Higgs. Improved limits on new physics come from lepton-flavor 
violation searches as in the slide on $\mu \rightarrow e \gamma$ shown by 
Francesco Renga \cite{renga} and from the ratio of leptonic two-body $K$ decays, 
$R_K = \Gamma(K \rightarrow e \nu)$ over $\Gamma(K \rightarrow \mu \nu)$ 
shown by Evgueni Goudzovski \cite{goudzovski}. Two of the most colorful plots 
are those showed by Vincenzo Chiochia where constraints on new physics come 
from top-quark-production asymmetries (forward-backward and charge asymmetries)
measured at the Tevatron and LHC \cite{chiochia} and by Nicola Serra showing 
limits from the rare processes $B \rightarrow \mu \mu$ and 
$B_S \rightarrow \mu \mu$ \cite{serra} taken from the presentation by David 
Straub at the EW Rencontres de Moriond this year. Finally, I would include 
in my postcard collection, the distributions of observed events in the search 
for $\tau \rightarrow \mu \mu \mu$ at LHCb shown by Paul Seyfert \cite{seyfert}. 

\subsection{Tension with the Standard Model}

Various fits to measurements of CKM unitarity triangles have been shown to 
highlight possible discrepancies in the single-phase paradigm of the Standard 
Model. Tensions with the Standard-Model overall fits have been observed, 
mention being made by Sebastien Descotes-Genon \cite{genon} and Koji Hara 
\cite{hara} of issues between the value of $\sin (2\beta)$ and the rate of $B$ 
decay to $\tau \nu$. Also mentioned by Sebastien were the semileptonic 
asymmetry in $B$-decay and $B_s$-decay parameters and the same-sign dimuon 
charge asymmetry measured by CDF and DZero at the Tevatron. Amarjit Soni showed 
a nice plot from a fit without inputs from semileptonic decays to address 
concerns over sensitivity to $V_{cb}$ \cite{soni}.  

Vera Luth noted that the ``tension" between inclusive and exclusive analyses of
semileptonic B decays remains \cite{luth}, while stated uncertainties on the 
branching fractions and on $|V_{ub}|$  and $|V_{cb}|$ are being reduced.
The search for discrepancies between measurements and Standard-Model predictions 
of $B$-decay rates has revealed a significant excess of events in 
$B \rightarrow D \tau \nu$ and in $B \rightarrow D^* \tau \nu$ (3.4 $\sigma$ 
when the two BaBar excesses are combined). This feature cannot be easily 
explained. Finally, I would point to the nice plots in the presentation by 
Rick Van Kooten \cite{vankooten} showing some tension in $B_S$ decay parameters.

\subsection{New signals and structures} 

Clearly, the observation of electron neutrinos coming from the oscillation of 
muon neutrinos, giving the large observed value of $\theta_{13}$, is a new 
signal. I show two more plots on this, which include the individual 
measurements from Daya Bay, Reno, Double Chooz, T2K, MINOS, KamLAND, and 
solar-neutrino measurements \cite{liu, litchfield}. 

We also saw unexpected structures in baryonic $B$ decays from BaBar as shown 
by Irina Nasteva \cite{nasteva} and, at 6.3 $\sigma$, the first observation of 
$B_S \rightarrow \phi \mu \mu$ from CDF shown by Rick Van 
Kooten \cite{vankooten}. 

The evidence for new $Z$-onium states is becoming more and more compelling, 
given the nice plots of $Z_{1,2} \rightarrow h_b(1,2) \pi$ from Belle shown 
by Jin Li \cite{li}. Another visually compelling set of plots demonstrated the 
suppression of the production the heavier upsilon mesons relative to the 
ground-state upsilon in heavy-ion collisions at CMS as shown by Zebo 
Tang \cite{tang}.

Another signal that is becoming clear with the recent increase of data at LHCb 
is the zero-crossing point in the forward-backward asymmetry in the decay of 
$B \rightarrow K^* \mu \mu$ shown by Nicola Serra \cite{serra}. I have also 
selected three slides from the presentation of BESIII results shown by 
Guangsung Huang \cite{huang}: the newly observed isospin-breaking decay of 
$\eta(1405) \rightarrow f_0(980) \pi^0$, an anomalous lineshape of the $f_0$ 
in the decay $J/\psi \rightarrow \gamma f_0 \pi^0$, and the first evidence of
$\psi(2S) \rightarrow \gamma \gamma J/\psi$. 

Perhaps the most unusual new signal was the first direct observation of 
time-reversal-symmetry violation in any system. The direct observation comes 
from using entangled $\Upsilon(4S)$ decays to tagged $B$ mesons. The results 
shown by Pablo Villanueva Perez \cite{perez} come from BaBar. 

As already noted, from a time-dependent Dalitz-plot analysis, we saw the 
first evidence of CPV in $B_S \rightarrow K_S K_S K_S$ in the talk by Giovanni 
Marchiori.

\subsection{Hints of new physics} 

Among the hints of new physics is the forward-back asymmetry in $t \bar{t}$ 
production at CDF and DZero, increasing with the mass of the $t \bar{t}$ system 
at CDF as shown by Marc Besancon \cite{besancon}. There are also hints of the 
Higgs at ATLAS and CMS shown by David Charlton \cite{charlton}, preceding the 
already-mentioned announcements after the FPCP conference. I continue to 
think of this as new physics, though most of you may already consider this a 
part of the Standard Model.

The surprisingly-large $D^0$ mixing observed has led to suggestions that 
there might be $CP$-symmetry violation in the charm sector.  Combining LHCb 
and CDF results on the difference of $CP$ asymmetries in $D^0$ decays to 
$K^-K^+$ and $\pi^-\pi^+$ led Vincenzo Vagnoni to say that the data ``is 
consistent with no $CP$ violation at 0.006\% CL" \cite{vagoni}. Even stated 
in this way, of course, at this point, we can only say that the 
data is inconsistent with no asymmetry at the given confidence level. The issue 
of $CP$ violation is still being debated among theorists! Can the asymmetry 
be due to Standard-Model long-distance effects?

\subsection{New techniques and looking beyond the lamp post} 

New techniques include both the application of new analysis methods and 
improvements made to previous analyses. Manuel Tobias Schiller showed LHCb 
multibody-decay results using combined Gronau-London-Wyler (GLW) and 
Atwood-Dunietz-Soni (ADS) methods and also taking advantage of the strong 
variation of hadronic parameters over the Daliz-plot phase space 
\cite{schiller}. David Charlton showed how analysis improvements allowed CDF 
to obtain more precise results in their search for the Higgs boson than what 
one would expect from the simple increase in data as the integrated luminosity 
grew over time \cite{charlton}. Similar improvements have been made at DZero, 
and we may expect similar things from the LHC experiments too, as they 
accumulate more data and experience. 

As mentioned earlier, searching for the unexpected is looking beyond the 
lamp post. Liang Sun showed the event distributions for searches for an 
unexpected $B^+ \rightarrow D^- \ell^+ \ell^+$ by Belle \cite{sun}.

\subsection{Postcards of the future}

I was going to reserve the last slide for postcard images of the new facilities 
shown in the first session of the final day. However, I decided to focus on the 
physics we expect most. The fact is, I am less certain what to expect than I 
have been for many years. So, I have left space in my postcard collection 
for next year's souvenirs!

\section{Summary}

As we flipped through the souvenir picture post cards of experiment results 
selected from all those presented at FPCP 2012, we have seen a wonderfully-rich 
abundance of new results. Yet, the selection presented here is necessarily 
incomplete. At best, it gives a sense of the impressive range of activity in 
flavor-physics and $CP$-violation experiments, and of the very high quality 
of the data and analyses being generated.

There were over 40 experiment talks! In these talks, there were presentations 
of results which included Standard-Model confirmations, significant reductions 
in uncertainty, New Physics spaces ruled out, new signals and structures, 
hints of New Physics, and still some tension with the Standard Model. And, 
there were also places where results were presented which explored the 
possibility of unexpected signals, looking beyond the lamp post.

There were, perhaps, fewer outstanding experimental issues relative to what was 
presented at FPCP 2011, less tension with the Standard Model this year. 
Nevertheless, there is growing disquiet over not seeing directions for the
answers to the questions about matter, energy, space, and time that were listed 
in my introduction. There is no certainty about the direction of the needed 
New Physics. [The signals around 125 $GeV/c^2$ shown by ATLAS and CMS after 
FPCP 2012 also seems consistent with the Standard-Model Higgs boson, though 
this is by no means yet proven.] We may hope for one or more bright new ideas 
that could provide additional lamp posts to light our way. But this 
hope is not something that we can count on soon. Thus, it is hard to predict 
what the post cards will look like next year at FPCP 2013 in Buzios, Brazil.
Nevertheless, given the huge data sets collected and anticipated, there is
excellent reason for hope! As has happened in the past, data may be the key
to future progress in our understanding.

\bigskip 
\begin{acknowledgments}

I close this last talk of the Flavor Physics and CP Violation 2012 meeting with 
a big and sincere thank you to all the presenters. Obviously, I have taken 
freely from their presentations for this summary. I also want to thank 
especially the organizers of this very enjoyable, interesting, and informative 
meeting.  

My work is supported, in part, by the US Department of Energy through Fermilab, 
which is operated by Fermi Research Alliance, LLC under Contract 
No. DE-AC02-07CH11359 with the Department of Energy.

\end{acknowledgments}

\bigskip 


\end{document}